\documentclass[aps,prl,reprint,superscriptaddress]{revtex4-1}

\usepackage{mathtools}
\usepackage{amsmath,amssymb}
\usepackage{color}
\usepackage{braket}
\usepackage{graphicx}
\usepackage{epstopdf}
\usepackage{float}
\usepackage{nccmath}
\usepackage[normalem]{ulem} 
\usepackage{textgreek}

\begin{document}


\title{Thermalization of light's orbital angular momentum in nonlinear multimode waveguide systems}


\author{Fan O. Wu}
\thanks{These two authors contributed equally to this work.}
\affiliation{CREOL/College of Optics and Photonics, University of Central Florida, Orlando, Florida 32816, USA}

\author{Qi  Zhong}
\thanks{These two authors contributed equally to this work.}
\affiliation{CREOL/College of Optics and Photonics, University of Central Florida, Orlando, Florida 32816, USA}

\author{Huizhong  Ren}
\affiliation{CREOL/College of Optics and Photonics, University of Central Florida, Orlando, Florida 32816, USA}

\author{Pawel S. Jung}
\affiliation{CREOL/College of Optics and Photonics, University of Central Florida, Orlando, Florida 32816, USA}
\affiliation{Faculty of Physics, Warsaw University of Technology, Koszykowa 75, 00-662 Warsaw, Poland}

\author{Konstantinos G. Makris}
\affiliation{Department of Physics, University of Crete, P.O. Box 2208, 71003 Heraklion, Greece}
\affiliation{Institute of Electronic Structure and Laser, Foundation for Research and Technology-Hellas, P.O. Box 1527, 71110 Heraklion, Greece}

\author{Demetrios N. Christodoulides}
\email[]{demetri@creol.ucf.edu}
\affiliation{CREOL/College of Optics and Photonics, University of Central Florida, Orlando, Florida 32816, USA}



\date{\today}

\begin{abstract}
We show that the orbital angular momentum (OAM) of a light field can be thermalized in a nonlinear cylindrical multimode optical waveguide. We find, that upon thermal equilibrium, the maximization of the optical entropy leads to a generalized Rayleigh-Jeans distribution that governs the power modal occupancies with respect to the discrete OAM charge numbers. This distribution is characterized by a temperature that is by nature different from that associated with the longitudinal electromagnetic momentum flow of the optical field. Counterintuitively and in contrast to previous results, we demonstrate that even under positive temperatures, the ground state of the fiber is not always the most populated in terms of power. Instead, because of OAM, the thermalization processes may favor higher order modes. A new equation of state is derived along with an extended Euler equation --- resulting from the extensivity of the entropy itself. By monitoring the nonlinear interaction between two multimoded optical wavefronts with opposite spins, we show that the exchange of angular momentum is dictated by the difference in OAM temperatures, in full accord with the second law of thermodynamics. The theoretical analysis presented here is corroborated by numerical simulations that take into account the complex nonlinear dynamics of hundreds of modes. Our results may pave the way towards high power optical sources with controllable orbital angular momenta, and at a more fundamental level, could shed light on the physics of other complex multimoded nonlinear bosonic systems that display additional conservation laws. 
\end{abstract}

\maketitle

Angular momentum plays a pivotal role in physics. In settings with a continuous rotational symmetry, Noether's theorem \cite{Thompson-AM,Cristina2020PRX} dictates that this quantity is conserved, an aspect that governs both the macroscopic and microscopic behaviors of a multitude of physical systems. These could range from the dynamics of spiral galaxies \cite{Genel2015AJ}, pulsars \cite{Oliveira2021AN} and neutron stars to the intriguing properties of quantum vortices in superfluids \cite{Boulier2016PRL} and superconductors. That fact that the electromagnetic field carries spin and/or orbital angular momentum (OAM) was recognized early on with the advent of Maxwell's electrodynamics \cite{BethPR1936,Jackson-CE}. Yet, it is only recently that the angular momentum of light was recognized as a new degree of freedom through which a wealth of opportunities could open up within the discipline of optics and photonics \cite{Yao2011AOP,Shen2019LSA,Molina2007NP,Ramos2020PRX}. In this respect, the orbital angular momentum of light has been systematically used in manipulating and guiding atoms \cite{He1995PRL}, in optical vortex soliton interactions \cite{Swartzlander1992PRL,Yuri1998PR}, in microscopies \cite{Xie2014PRL}, optical ablation \cite{Toyoda2013PRL}, quantum entanglement \cite{Sit2017Optica,Stav2018S,Erhard2018LSA}, astronomy \cite{Harwit2003AJ,Lee2006PRL,Swartzlander2008OE,Tamburini2011NP} and high-speed communication systems \cite{Gibson2004OE,Wang2012NPho}, to mention a few.  Similar ideas have lately permeated the area of statistical optics as well as the new emerging field of optical metasurfaces where new methodologies have been developed to generate and detect OAM light states \cite{Kildishev2013S,Sounas2013NC,Karimi2014LSA,Maguid2016S,Devlin2017S}.  

On the other hand, the way angular momentum appears in statistical mechanics is more subtle. As indicated by E. T. Jaynes and  H. B. Callen \cite{Jaynes1957PR,Callen-Thermodynamics}, if the OAM $\vec{L}$ is a constant of the motion, it can then assume the role of an extensive variable --- in addition to that of the energy. As such, it must be paired with an intensive quantity or a generalized force $\vec{q}$ that acts as a ``temperature" in the canonical distribution function. Under these conditions, one should expect that the entropy of the system must also be extensive with the respect to the OAM, thus satisfying a key postulate of thermodynamics. While the idea of direct OAM thermalization in black hole environments has been discussed in the past \cite{Bardeen1973CMP,Bekenstein1980PT,Carlip2014IJMPD}, the prospect of observing similar phenomena in low energy settings is rather rare. In general, as opposed to energy conservation that is always guaranteed in an isolated system because of time translation symmetry, the OAM happens to be more demanding. At this point, the question naturally arises as to whether one can identify physical systems where the OAM (or the OAM ``quantum" numbers) can explicitly govern thermodynamic processes very much like its counterpart, the energy. If so, would that be possible in such a scenario to induce irreversible OAM exchange as dictated by the second law of thermodynamics?

\begin{figure}[!t]
	\includegraphics[width=\linewidth]{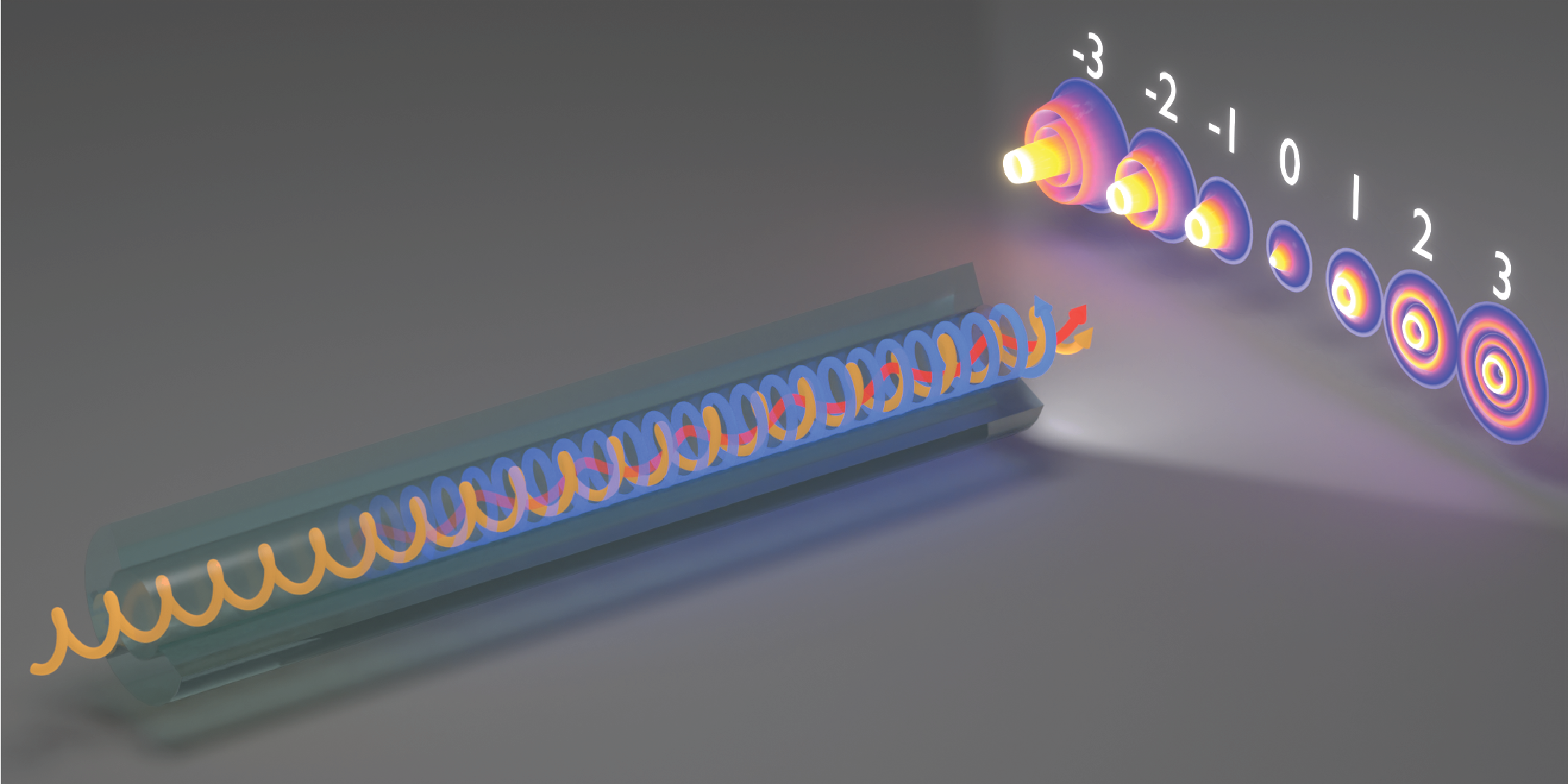}
	\caption{The OAM of light propagating in a multimode cylindrical structure can be thermalized because of nonlinearity in a way that favors either positive or negative OAM numbers. These are irreversible processes that result from the maximization of the optical entropy $S$ of the system.}
	\label{Fig1}
\end{figure}

In this Letter, we show for the first time, that the OAM of a light beam propagating in a nonlinear cylindrical waveguide structure can be thermalized in a manner that directly depends on the discrete OAM number of a particular mode. Along these lines, we demonstrate that the optical partition function of this arrangement is no longer separable with respect to the conservation laws of longitudinal momentum and OAM. 
 Even more importantly, upon attaining thermal equilibrium, the OAM is now governed by its own optical temperature $T_L$ --- a temperature that has nothing to do with that assigned to its energy counterpart $T_U$ (Fig. \ref{Fig1}). In this respect, a generalized Rayleigh-Jeans distribution can be derived, that even under positive temperature ($T_U$) conditions, does not always favor the ground state. New equations of state are obtained that relate the extensive variables with their corresponding generalized forces. Finally, based on entropic principles, we show that the OAM temperature $T_L$ is a true thermodynamic quantity in the sense that it dictates the OAM exchange between two optical beams of opposite spin when traversing a nonlinear medium.  The predicted theoretical results are in excellent agreement with numerical simulations.

To this end, let us consider a weakly guiding, nonlinear cylindrical waveguide having a normalized refractive index distribution $V(r)$, where $r$ represents the cylindrical radial coordinate. In Kerr nonlinear media, the optical field  $u(x,y,z)$ evolution along the $z$ direction is governed by the normalized nonlinear Schr\"odinger equation: 
\begin{equation}\label{Eq:NLS}
	i \frac{\partial u}{\partial z} + \left(\frac{\partial^2 u}{\partial x^2}+\frac{\partial^2 u}{\partial y^2}\right)+ V(x,y) u +|u|^2u=0.
\end{equation}
In this system, the field orbital angular momentum is given by $\vec{L}=\iint_{-\infty}^{\infty} \vec{r} \times \vec{p} dxdy$, where $\vec{p}=\frac{i}{2} (u\nabla_{\perp} u^* - u^* \nabla_{\perp} u)$ is the transverse momentum density, and $\nabla_{\perp}=\hat{x} \frac{\partial}{\partial x} + \hat{y} \frac{\partial}{\partial y}$. Given that $\vec{L} \parallel \hat{z}$, from this point on, $L$ will be treated as a scalar quantity.  The conservation laws $C$ associated with Eq. (\ref{Eq:NLS}) are known to satisfy $\{C,H\}_{u,\pi}=\iint_{-\infty}^{\infty} \left(\frac{\delta \mathcal{C}}{\delta u}  \frac{\delta \mathcal{H}}{\delta \pi}-\frac{\delta \mathcal{H}}{\delta u}  \frac{\delta \mathcal{C}}{\delta \pi} \right)dxdy=0$, where $\mathcal{H}=-i (u_x \pi_x+u_y \pi_y)+iVu\pi + \frac{1}{2} u^2 \pi^2$ is the system's Hamiltonian density, $\pi=iu^*$ is the canonical  field momentum, $\{\cdot\}$ denotes a Poisson bracket and $\delta$ represent functional derivatives.  In addition to the total power $\mathcal{P}=\iint_{-\infty}^{\infty} |u|^2 dxdy$ that remains constant throughout propagation, one can show that in a fully cylindrical structure [$V(x,y)=V(r)$], the field orbital angular momentum $L$ is also an invariant in this nonlinear system (see Supplemental Material  \cite{SM}). 

In general, an optical field propagating in the aforementioned cylindrical waveguide (supporting $M$ modes) can be represented as a superposition of its underlying bound states $\psi_n (r,\phi,z)=R_n(r) e^{i \varepsilon_n z} e^{i l_n \phi}$, i.e., $u=\sum_{n=1}^M c_n \psi_n$. Here, $\varepsilon_n$ stands for the propagation constant or eigenvalue of the eigenmode $\psi_n$, $l_n\in \{0,\pm 1,\pm 2,\dots\}$ denotes its discrete OAM charge, $R_n (r)$ is the radial part of the eigenfunction while the complex number $c_n(z)$ describes the power occupancy ($|c_n |^2$) of this state during propagation.  In this representation the power and OAM invariants can be written as $\mathcal{P}=\sum_{n=1}^M |c_n |^2$  and $L=\sum_{n=1}^M l_n |c_n |^2$  \cite{SM}. On the other hand, the Hamiltonian constant of the motion $H=\iint \mathcal{H} dxdy$ involves both a linear ($-U$) and a nonlinear component ($H_\text{NL}$), that is $H=-U+H_\text{NL}$ \cite{Wu2019NPho}. Given that the system is operating in the weakly nonlinear regime, one finds that the Hamiltonian is dominated by its linear contribution, and as a result the quantity $U=-\sum_{n=1}^M \varepsilon_n |c_n |^2$  now assumes the role of the third invariant. Interestingly, this ``internal optical energy" $U$ is associated with the Minkowski longitudinal electromagnetic momentum flowing in this guiding arrangement \cite{Haus1976JOSA}. 

The existence of these three invariants $(\mathcal{P},U,L)$ in this weakly nonlinear heavily multimode optical fiber now allows one to deploy principles from statistical mechanics. What facilitates this approach is the absence of required conservation laws ($M-3$), an aspect that leads to chaotic behavior in the evolution of the modal occupancies $|c_n (z)|^2$ and therefore allows the system to ergodically explore its phase space. In other words, all possible microstates $(c_1,c_2,...,c_M)$ that lie on the manifolds of constant angular momentum ($L$), power ($\mathcal{P}$) and energy ($U$) are accessed with equal probability because of nonlinearity. To derive the state functions that relate the thermodynamically extensive variables $(\mathcal{P},U,L,M)$ to the entropy $S$, we adopt a grand canonical description in a phase space constructed by $J_n\equiv|c_n |^2$ $(n=1,2,…,M)$ \cite{Makris2020OL}. In this grand canonical frame, the local state of the system is described by a normalized probability density distribution $\rho(J_1,…,J_M)$ where  $\int_0^{\infty} \rho(J_1,...J_M) \prod_{n=1}^M dJ_n=1$. Once the system attains thermal equilibrium, the average values of $\mathcal{P}$, $L$ and $U$ serve as the invariants of this grand canonical system. In turn, the probability density $\rho$ can be used to construct the Gibbs entropy 
\begin{equation}
	S=-\int_0^{\infty} \rho(J_1, ..., J_M) \ln \rho(J_1, ..., J_M) \prod\limits_{n=1}^{M} dJ_n,
\end{equation}
which can be maximized by means of Lagrange multipliers, i.e., 
\begin{equation}
	\rho(J_1,...,J_M)=\frac{e^{-\alpha \mathcal{P}(J_1,...,J_M) - \beta U(J_1,...,J_M) - \gamma L(J_1,...,J_M)}}{\mathcal{Z}},
\end{equation}
where $\alpha$, $\beta$ and $\gamma$ represent constants associated with the three invariants. In this case, the generalized grand partition function is given by \cite{SM}
\begin{subequations}\label{Eq:Z}
	\begin{align}
		\mathcal{Z}
		&=\int_0^{\infty} e^{-\alpha \sum\limits_{n=1}^M J_n + \beta \sum\limits_{n=1}^M \varepsilon_n J_n - \gamma \sum\limits_{n=1}^M l_n J_n}  \prod\limits_{n=1}^{M} dJ_n  \\ 
		&=\prod\limits_{n=1}^M \frac{1}{-\alpha + \beta \varepsilon_n - \gamma l_n}.
	\end{align}
\end{subequations}
At equilibrium, the mean value $\left\langle J_m \right\rangle=\left\langle |c_m|^2 \right\rangle=\int_0^{\infty}\rho(J_1,...,J_M)J_m\prod_{n=1}^M dJ_n$ can be directly obtained from the generalized partition function $\mathcal{Z}$ through the relations: $\left\langle J_m\right\rangle=\frac{1}{\beta}\frac{\partial\ln{\mathcal{Z}}}{\partial\varepsilon_m}=-\frac{1}{\gamma}\frac{\partial\ln{\mathcal{Z}}}{\partial l_m}$ \cite{SM}. From here, one can find that $\left\langle J_m\right\rangle=\frac{-1}{\alpha-\beta\varepsilon_m+\gamma l_m}$. We then introduce two intensive quantities by defining  $\beta\equiv-1/T_U$ and $\gamma\equiv -1/T_L$, i.e., an energy temperature $T_U$ and an OAM temperature $T_L$, that are conjugate to the extensive variables $U$ and $L$, respectively. Here,  $\alpha$ is another intensive quantity that is conjugate to $\mathcal{P}$, which serves as the chemical potential of the system. As a result, upon thermalization, the average power occupancy of each mode is found to obey a generalized Rayleigh-Jeans distribution 
\begin{equation}\label{Eq:RJ}
	\left\langle|c_n|^2\right\rangle=\frac{-1}{\alpha+\frac{\varepsilon_n}{T_U}-\frac{l_n}{T_L}}.
\end{equation}

The generalized Rayleigh-Jeans distribution in Eq. (5) resulting from a generalized Gibbs ensemble \cite{Langen2015S,Cassidy2011PRL} is central in this work. It implies that not only the internal energy $U$ can be thermalized (via eigenvalue $\varepsilon_n$), but also the OAM since it explicitly involves the topological charge of the mode $l_n$ --- an aspect that has a profound effect on the modal power distributions (Fig. \ref{Fig1}). Because of OAM, the possibility exists that the ground state (the lowest order mode) is no longer the most populated in terms of power. Even more importantly, $T_L$ is a true thermodynamic quantity that governs the flow of OAM between two optical beams. From this generalized distribution, the following global equation of state can be derived  \cite{SM}
\begin{equation}\label{Eq:StateFunction1}
	-\alpha \mathcal{P} +\frac{U}{T_U} +\frac{L}{T_L} = M,
\end{equation}
which relates the three intensive variables $\alpha$, $T_U$ and $T_L$  to the four extensive quantities $\mathcal{P}$, $U$, $L$ and $M$. 

Based on these premises, one can formally show that the entropy  $S$ of this optical multimode system can be directly expressed in terms of the modal occupancies via $S=\sum_{n=1}^M \ln (\left\langle|c_n |^2 \right\rangle)$ \cite{Wu2019NPho,Haus1976JOSA,Makris2020OL,SM}. In addition, the fundamental equation of thermodynamics demands that $S=S(\mathcal{P},U,L,M)$. Hence, two temperatures can be entropically defined through $T_U^{-1}\equiv \partial S/\partial U$, and  $T_L^{-1}\equiv \partial S/\partial L$. Similarly, a generalized chemical potential $\alpha$ and an optical thermodynamic pressure  $\tilde{p}$ can also be introduced using $\alpha \equiv -\partial S/\partial \mathcal{P}$ and $\tilde{p} \equiv \partial S/\partial M$  \cite{SM}. From here, a corresponding Euler equation can be obtained which is a direct manifestation of the extensivity of the entropy itself with respect to $(\mathcal{P},U,L,M)$,
\begin{equation}\label{Eq:StateFunction2}
	S=-\alpha \mathcal{P}+\frac{U}{T_U} +\frac{L}{T_L} +\tilde{p} M. 
\end{equation}

\begin{figure}[!t]
	\includegraphics[width=3.4in]{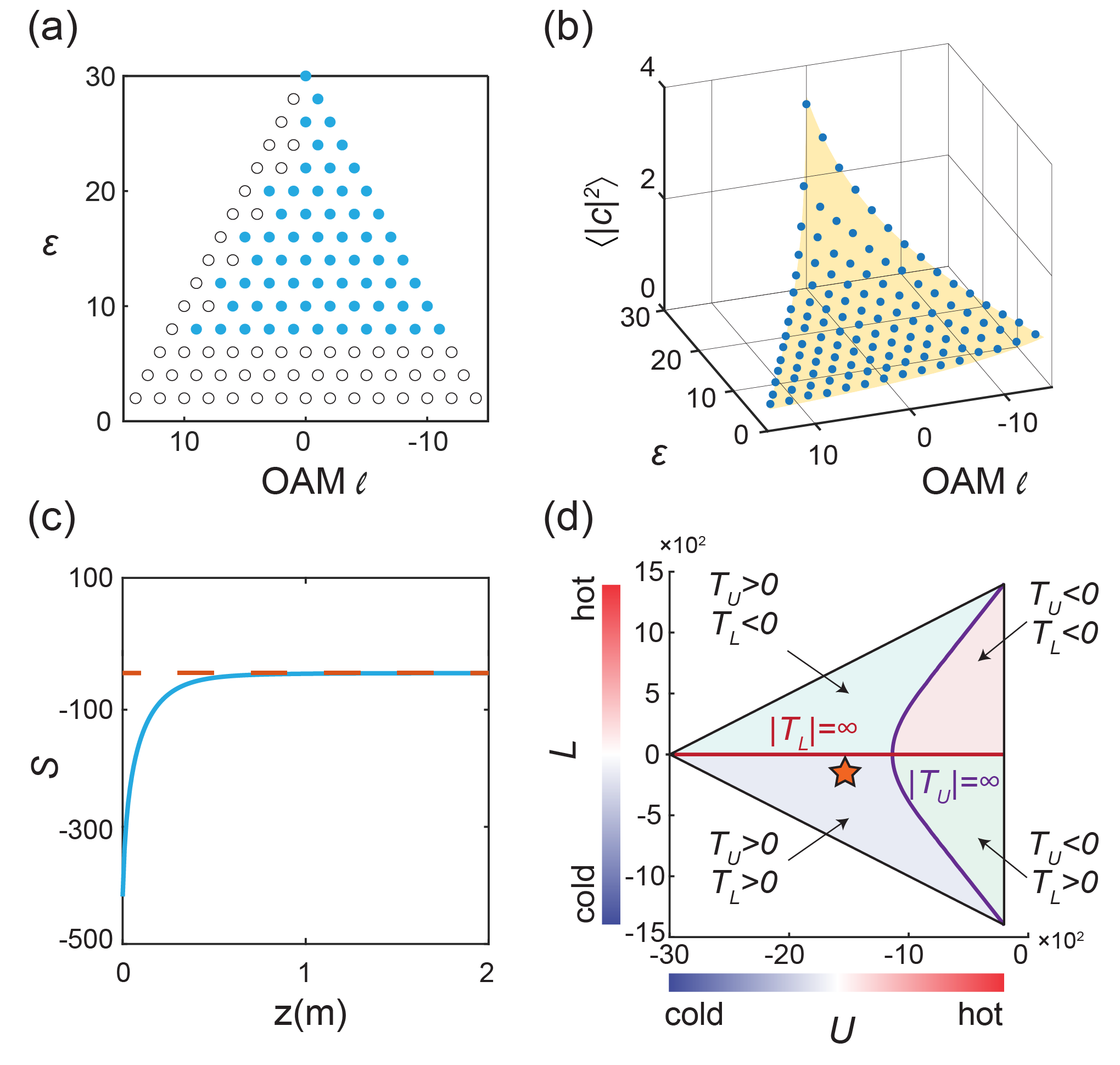}
	\caption{Thermalization dynamics of the OAM associated with an optical beam propagating in a nonlinear parabolic graded-index fiber. (a) Initially excited modes (blue dots) are plotted on the $\varepsilon - l$ triangular eigenvalue grid of a parabolic fiber. All these modes are equally excited. (b) The theoretically predicted modal occupancies $\left\langle|c_n |^2\right\rangle$ lying on the yellow surface are in excellent agreement with those obtained from numerical simulations (blue dots).  (c) During propagation, the optical entropy monotonically increases (blue curve) until it reaches its maximum value, predicted from Eq. (\ref{Eq:RJ}). (d) Depending on initial conditions $(U,L)$, the system can settle into four different temperature regimes which are separated by the infinite temperature curves of energy (red) and OAM (purple). The red star denotes the initial $(U,L)$ quantities that lead to (b) with $T_U>0$, $T_L>0$. This diagram was obtained for $\mathcal{P}=100$.}
	\label{Fig2}
\end{figure}

In what follows, we corroborate our theoretical formalism by performing a series of numerical simulations in nonlinear parabolic and step-index optical fibers. In this respect, the results presented in Eqs. (\ref{Eq:RJ}) and (\ref{Eq:StateFunction1}) will be employed to predict the OAM thermalization once equilibrium conditions are attained. As a first example, let us consider a parabolic silica fiber having $M=120$ modes, conveying in total 100 kW of power at a wavelength of 1064 nm \cite{SM}. For this case, 62 modes of this fiber are evenly excited [Fig. \ref{Fig2}(a)]. In normalized units, these launching conditions correspond to $\mathcal{P}=100$, $U=-1468$ and $L=-140$. For this scenario, our theory predicts that at thermal equilibrium, the propagating optical wavefront is expected to settle into an OAM temperature $T_L=14.5$, an energy temperature $T_U=15.2$ and a generalized chemical potential $\alpha=-2.26$. To verify these predictions, we numerically solved  Eq. (\ref{Eq:NLS}) under the same initial conditions. For each run, we allowed the phases of the input $c_n$ coefficients to statistically vary in order to construct the statistical ensembles. Figures. \ref{Fig2}(b) and (c) show that an excellent agreement exists between the theoretically anticipated results and the numerical computations after OAM thermal equilibrium is reached at $\sim$1 m of propagation. In this case, the modal occupancies $\left\langle |c_n|^2 \right\rangle$  are displayed on a two-dimensional triangular map that is specified by the accessible normalized propagation constants $\varepsilon_n$ and OAM numbers $l_n$ associated with the optical modes. For these initial conditions, the energy temperature $T_U$ is positive and therefore the fundamental mode (LP$_{01}$) is the one mostly populated. Meanwhile, the OAM temperature $T_L$ so happens to be negative, in which case the optical power resides mostly in the right-handed rotating modes with positive OAM charges ($l_n>0$). It is important to note that, in all cases, the initial conditions uniquely determine the  equilibrium intensive variables $(T_L,T_U,\alpha)$ of this system. As Fig. \ref{Fig2}(d) shows, both positive and negative $T_L$ temperatures are possible, favoring left-handed and right-handed OAM groups of modes, respectively. Given that the OAM numbers in a cylindrical structure always come in pairs ($\pm l$), if the input OAM is zero, the system will relax into an infinite $T_L$ temperature [Fig. \ref{Fig2}(d)], which leads to OAM equipartition and hence the Rayleigh-Jeans distribution of Eq. (\ref{Eq:RJ}) assumes a standard form \cite{Dyachenko1992PD}.

\begin{figure}[!t]
	\includegraphics[width=3.4in]{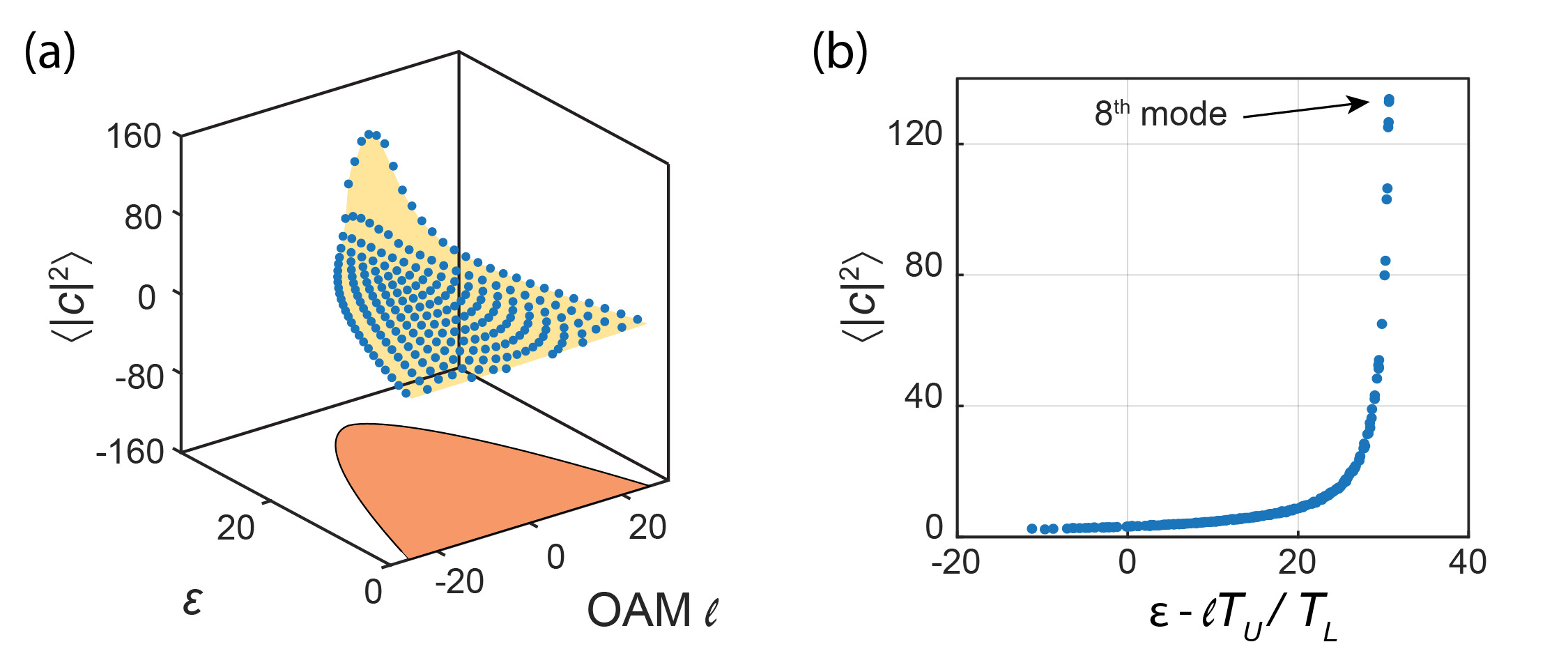}
	\caption{Thermalization OAM dynamics associated with an optical beam propagating in a nonlinear step-index optical fiber. (a) The theoretically predicted modal occupancies $\left\langle|c_n |^2\right\rangle$ lying on the yellow surface are in good agreement with those obtained from numerical simulations (blue dots). For this example ($\mathcal{P}=3200$), the OAM temperature is negative while the energy temperature is positive. Note that the $\varepsilon-l$ diagram (orange area) of a step-index fiber has now a parabolic-like shape. (b) Modal occupancies as a function of $(\varepsilon-l T_U/T_L)$. For this particular example, $T_U/T_L=-0.5$ and as a result, the lowest order mode, the ground state, is no longer the most favored.  Instead, the higher-order LP$_{+3,1}$ mode now conveys most of the power.  }
	\label{Fig3}
\end{figure}

Our theory is universal in the sense that it can be deployed to predict OAM thermalization in any cylindrical multimode nonlinear structure. Figure \ref{Fig3}(a) shows the results of OAM thermalization in a step-index fiber having $M=234$ modes. For the step-index case, the $\varepsilon-l$ map now assumes a parabolic-like shape. For the excitation conditions used in this simulation,  $\mathcal{P}=3200$, $U=-75420$ and the $L=8776$, for which values our theory predicts $T_U=104$, $T_L=-204$ and $\alpha=-0.04$. As Figs. \ref{Fig3}(a) and (b) show, there is an excellent agreement between the numerical simulations and the theoretically anticipated results.  Counterintuitively, in this case, while the energy temperature $T_U$ is positive, the fundamental mode (LP$_{01}$) is no longer the most populated.  Instead, the 8$^\text{th}$  mode LP$_{+3,1}$ ($l=+3$) carries most of the power. This anomaly in the distribution results from the interplay between the more convolved $\varepsilon-l$ map that corresponds to a step-index fiber and the non-separability of $U$ and $L$ in the generalized Rayleigh-Jeans distribution. To intuitively understand this behavior, one can assign effective eigen-energies in this map according to $\varepsilon_{n,\text{eff}}=\varepsilon_n- \l_n T_U/T_L$ that angularly [$\theta=\arctan(T_U/T_L )$] project this data into a standard Rayleigh-Jeans curve. This is illustrated in Fig. \ref{Fig3}(b) where the 8$^\text{th}$ mode is the first in line to assume the role of an ``effective ground state". In this respect, the ratio of the two temperatures $T_U/T_L$ can be used to select the highest populated mode in this system. 

\begin{figure}[!t]
	\includegraphics[width=3.4in]{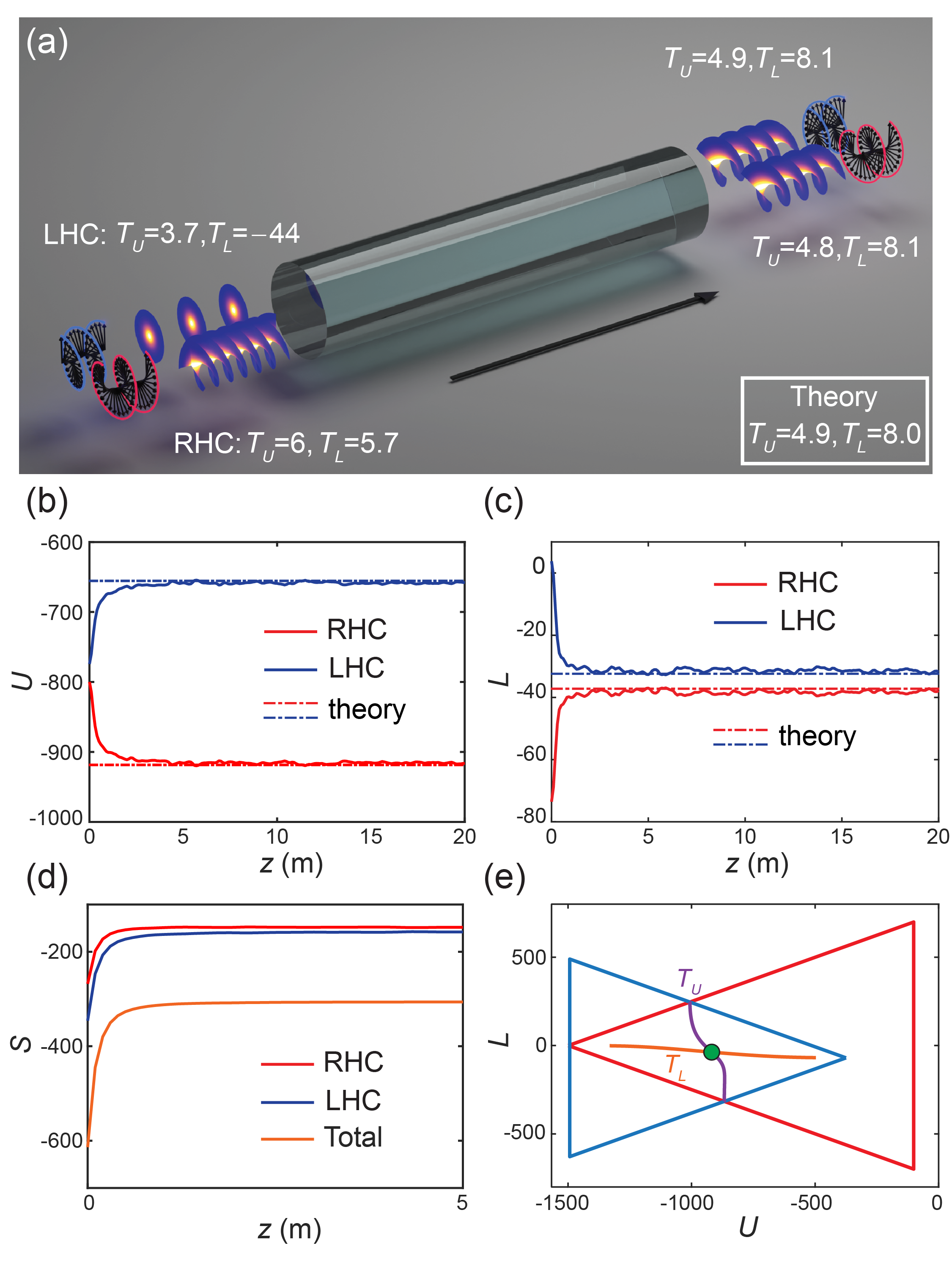}
	\caption{(a) When a RHC beam conveying significant negative OAM (red) and a LHC beam having almost 0 OAM (blue) are launched into the same fiber, they eventually settle into the same energy temperature $T_U \approx 4.9$ and OAM temperature $T_L \approx 8.0$. During propagation, the longitudinal momentum $U$ is transferred from the RHC to LHC polarization (b), whereas the OAM flows in the opposite direction given that initially $T_{U,u}>T_{U,v}>0$ and $T_{L,v}<0<T_{L,u}$ (negative temperatures are hotter than positive). (d) Meanwhile, the total optical entropy monotonically increases until it stabilizes as demanded by the second law of thermodynamics. (e) The final state can be uniquely predicted from the initial conditions $\mathcal{P}_u$, $\mathcal{P}_v$, $U_\text{total}$ and $L_\text{total}$, through the intersection of the $T_U$, $T_L$ manifolds in the $(U,L)$ plane \cite{SM}. }
	\label{Fig4}
\end{figure}

Next, we show that the OAM temperature $T_L$ is an actual thermodynamic force that governs the direction of OAM exchange between two subsystems, just like its energy counterpart. Figure \ref{Fig4}(a) illustrates a scenario where two circularly polarized beams are launched in the same nonlinear parabolic silica fiber. In this case, the total electric field  with components $\ket{R}$ and $\ket{L}$ can be written in terms of slowly varying envelopes $(u,v)$ as $\vec{E} =u \ket{R}+ v \ket{L}$, whose evolution equations can be found in \cite{SM}. In this arrangement, one can formally identify the following conservation laws \cite{SM}: $\mathcal{P}_u=\text{const.}$, $\mathcal{P}_v=\text{const.}$, $L_\text{total}=L_u+L_v=\text{const.}$, $U_\text{total}=U_u+U_v=\text{const.}$. In the example that follows the right-hand circularly (RHC) polarized light (red) is hotter in energy temperature $T_U$, while the left-handed (LHC, blue) is hotter in terms of its OAM. Moreover, at the input $L_u(0)=-74$, $L_v(0)=4 \ (L_\text{total}=-70)$, $\mathcal{P}_u=50$, $\mathcal{P}_v=40$, $U_u(0)=-800$, $U_v(0)=-774 \ (U_\text{total}=-1574)$. While the two circular polarizations do not exchange power $\mathcal{P}$, they can exchange energy $U$ and OAM $L$ according to the second law of thermodynamics 
\begin{equation}
	dS_T=(\frac{1}{T_{U,u}} -\frac{1}{T_{U,v} })dU_u+(\frac{1}{T_{L,u}} -\frac{1}{T_{L,v} })dL_u \geq0
\end{equation}
given that $dU_u=-dU_v$ and $dL_u=-dL_v$. As a result, during co-propagation, both energy and OAM are exchanged until the two polarizations settle into the same temperatures $(T_{U,u}=T_{U,v}, T_{L,u}=T_{L,v})$, as expected from actual thermodynamic forces. Figures \ref{Fig4}(b)--(d) demonstrate this behavior. Note that the equilibrium temperatures and chemical potentials can be directly predicted [Fig. \ref{Fig4}(e)] from the theoretical formalism presented above. 

In conclusion, we have shown that the OAM of a light beam propagating in a nonlinear cylindrical waveguide structure can be thermalized. A generalized Rayleigh-Jeans distribution has been derived that explicitly depends on the OAM charge. The interplay between energy and OAM can lead to an anomaly in the power distributions among modes, thus allowing higher-order modes to be favored. The exchange of OAM between two subsystems formally indicates that the OAM temperature acts like a true thermodynamic force.


\begin{acknowledgments}
This work was partially supported by ONR MURI (N00014-20-1-2789), AFOSR MURI (FA9550-20-1-0322, FA9550-21-1-0202), National Science Foundation (NSF) (DMR-1420620, EECS-1711230), MPS Simons collaboration (Simons grant 733682), W. M. Keck Foundation, US–Israel Binational Science Foundation (BSF: 2016381), US Air Force Research Laboratory (FA86511820019), the Polish Ministry of Science and Higher Education (1654/MOB/V/2017/0) and the Qatar National Research Fund (grant NPRP13S0121-200126).
\end{acknowledgments}

%

\end{document}